\documentclass[journal=jpcl,manuscript=letter]{achemso}

\usepackage{chemformula} 
\usepackage[T1]{fontenc} 
\usepackage{graphicx}

\author{Yiying Yan}
\affiliation{Department of Physics, School of Science, Zhejiang University of Science and Technology, Hangzhou 310023, China}
\email{yiyingyan@zust.edu.cn}
\author{Tadele T. Ergogo}
\affiliation{Department of Physics, School of Science, Zhejiang University of Science and Technology, Hangzhou 310023, China}
\author{Zhiguo L\"{u}}
\affiliation{Key Laboratory of Artificial Structures and Quantum Control
(Ministry of Education), School of Physics and Astronomy,
Shanghai Jiao Tong University, Shanghai 200240, China}
\author{Lipeng Chen}
\affiliation{Max Planck Institute for the Physics of Complex Systems, N\"{o}thnitzer Str., 38, 01187 Dresden, Germany}
\author{JunYan Luo}
\affiliation{Department of Physics, School of Science, Zhejiang University of Science and Technology, Hangzhou 310023, China}
\author{Yang Zhao}
\affiliation{Division of Materials Science, Nanyang Technological University, Singapore 639798, Singapore}\email{YZhao@ntu.edu.sg}

\title{Lamb Shift and the Vacuum Rabi Splitting in a Strongly Dissipative Environment}

\begin{document}



\begin{abstract}
We study the vacuum Rabi splitting of a qubit ultrastrongly coupled to a high-$Q$ cavity mode and a radiation reservoir. Three methods are employed: a numerically exact variational approach with a multiple Davydov ansatz, the rotating-wave approximation (RWA), and the transformed RWA. Agreement between the variational results and the transformed RWA results is found in the regime of validity of the latter, where the RWA breaks down completely. We illustrate that the Lamb shift plays an essential role in modifying the vacuum Rabi splitting in the ultrastrong coupling regime, leading to off-resonant qubit-cavity coupling even though the cavity frequency equals to the bare transition frequency of the qubit. Specifically, the emission spectrum exhibits one broad low-frequency peak and one narrow high-frequency peak in the presence of relatively weak cavity-qubit coupling. As the cavity-qubit coupling increases, the low-frequency peak narrows while the high-frequency peak broadens until they have similar widths.

\end{abstract}

\section{TOC Graphic}
\includegraphics{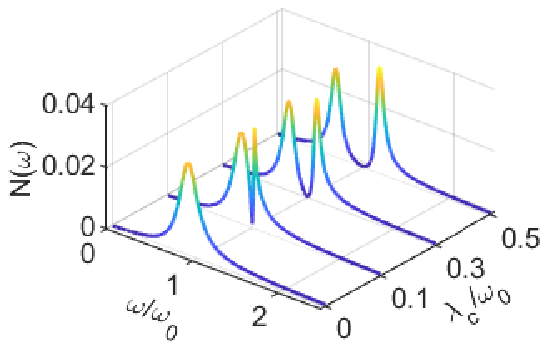}



Artificial atoms such as superconducting circuit qubits can be used as a paradigm for studying light-matter interaction in quantum optics in the ultrastrong coupling regime ~\cite{Forn_D_az_2010,Yoshihara_2016,Yoshihara_2017,Forn_D_az_2019,Frisk_Kockum_2019,Blais_2021,Thomas_2021}, where the coupling strength is comparable with the transition frequency of the system. In the ultrastrong coupling regime, the rotating-wave approximation (RWA) breaks down and the counter-rotating coupling cannot be neglected.
Many studies have shown a wide variety of effects of the counter-rotating coupling, such as the generation of entangled state~\cite{Ashhab_2010}, the virtual-photon-dressed ground state~\cite{Cirio_2016,De_Liberato_2017}, and Bloch-Siegert shift~\cite{Forn_D_az_2010,Wang_2020}. On the other hand, the ultrastrong light-matter coupling leads to the formation of polaritons, highly hybridized light-matter states, which are closely relevant for the control of chemical reactions~\cite{Herrera_2016,Mart_nez_Mart_nez_2017,Sidler_2020,Cederbaum_2021}.

One of the simplest models of light-matter coupling, the quantum Rabi model describes the interaction between a two-level system (qubit) and a single cavity mode~\cite{scully,Braak_2011}, and the vacuum Rabi splitting can occur when the coupling strength between the cavity and two-level emitter is much larger than the spontaneous emission rate of the latter, whereby the emission spectrum exhibits two separated Lorentzian peaks~\cite{Thompson_1992,Auff_ves_2008}. This phenomenon has been observed in the context of the atoms~\cite{Thompson_1992}, semiconductor quantum dots~\cite{Reithmaier_2004,Yoshie_2004}, and superconducting circuit qubit~\cite{Wallraff_2004}. Furthermore, the measured spectra are qualitatively explained by the RWA theory, which predicts a symmetric two-peaked spectrum if the qubit is resonantly coupled to the cavity mode.
With the counter-rotating coupling between the qubit and the cavity taken into account, the spectrum is found to be asymmetric~\cite{Cao_2011}. To the best of our knowledge, most of the studies on the vacuum Rabi splitting focus on the strong or ultrastrong coupling between the qubit and the cavity in the presence of the weak dissipation of the qubit. Few efforts have been devoted to the regimes where the qubit is ultrastrongly coupled to both the cavity and reservoir, which is realizable in the superconducting circuits~\cite{Forn_D_az_2016}.

As is well known, the ultrastrong coupling between the qubit and the reservoir gives rise to not only the non-Markovian nature of highly dissipative dynamics but also a significant Lamb shift, i.e., the transition frequency of the qubit will be renormalized from its bare counterpart~\cite{Leggett_1987,D_az_Camacho_2016}. Moreover, the magnitude of such frequency renormalization is of the order that is nonnegligible in comparison with the bare transition frequency of the qubit~\cite{Forn_D_az_2016}. Consequently, it may be expected that the Lamb shift plays an important role in the spontaneous emission. In addition,
the coupling can be engineered sufficiently strong such
that the spontaneous emission rate of the qubit becomes comparable or even exceeds its transition frequency.
It remains unexplored that the influence of the Lamb shift and strong dissipation on the vacuum Rabi splitting.

In this work, we employ the Dirac-Frenkel time-dependent variational principle~\cite{frenkel} with the multiple Davydov $D_1$ (multi-$D_1$) ansatz~\cite{Yan_2020,Wang_2016,Deng_2016} and two approximate analytical methods to study the vacuum Rabi splitting in the spontaneous emission spectrum of a qubit ultrastrongly coupled to a high-$Q$ cavity and a radiation reservoir, in which the spontaneous emission rate of the qubit is comparable with its transition frequency. One of our analytical methods is based a unitary transformation and resolvent formalism, which is referred to as the transformed RWA (TRWA) treatment. The other is based on the widely used RWA. Excellent agreement between the variational approach and TRWA is found in the regime of validity of the latter, while the RWA completely breaks down. We find that the Lamb shift plays an essential role in the vacuum Rabi splitting in the ultrastrong coupling regime, which leads to off-resonant qubit-cavity coupling in spite of the fact that the cavity frequency is equal to the bare transition frequency of the qubit.

We consider that a qubit is coupled to a high-$Q$ cavity mode and
a radiation reservoir. In the high-$Q$ limit, the dissipation of the cavity is assumed to be much weaker than that of the qubit and thus is neglected. The Hamiltonian of the
total system is given by~\cite{Henriet_2014} ($\hbar=1$)

\begin{eqnarray}
H & = & H_{{\rm Rabi}}+H_{{\rm R}}+H_{{\rm I}},\label{eq:Ham}
\end{eqnarray}
\begin{equation}
H_{{\rm Rabi}}=\frac{1}{2}\omega_{0}\sigma_{z}+\omega_{c}b_{c}^{\dagger}b_{c}+\frac{1}{2}\lambda_{c}\sigma_{x}(b_{c}^{\dagger}+b_{c}),
\end{equation}
\begin{equation}
H_{{\rm R}}=\sum_{k}\omega_{k}b_{k}^{\dagger}b_{k},
\end{equation}
\begin{equation}
H_{{\rm I}}=\frac{1}{2}\sigma_{x}\sum_{k}\lambda_{k}(b_{k}^{\dagger}+b_{k}).
\end{equation}
$H_{{\rm Rabi}}$ describes the interaction between the qubit and the cavity mode. $H_{{\rm R}}$ is the free Hamiltonian
of the reservoir and $H_{{\rm I}}$ describes the interaction between
the qubit and reservoir. The above notations are defined
as follows. $\omega_{0}$ is the bare transition frequency of the qubit and $\sigma_{i}$ ($i=x,y,z$) is the Pauli matrix. $\omega_{c}$
and $b_{c}$ ($b_{c}^{\dagger}$) are the frequency and annihilation
(creation) operator of the cavity mode, respectively. $\omega_{k}$
and $b_{k}$ ($b_{k}^{\dagger}$) are frequency and annihilation (creation) operator
of the $k$th mode of the reservoir, respectively. $\lambda_{c}$
and $\lambda_{k}$ are the coupling constants. The
interaction between the qubit and the reservoir is characterized
by the Ohmic spectral density function
\begin{equation}
J(\omega)=\sum_{k}\lambda_{k}^{2}\delta(\omega_{k}-\omega)=2\alpha\omega\exp(-\omega/\omega_{{\rm cut}}),
\end{equation}
where $\alpha$ is dimensionless coupling strength and $\omega_{{\rm cut}}$
is the cut-off frequency.

We calculate the dynamics and the emission spectrum by using the Dirac-Frenkel time-dependent variational principle
and the multi-$D_{1}$ trial state. The Dirac-Frenkel variational
principle states that the optimal solutions to the time-dependent
Schr\"{o}dinger equation can be obtained via~\cite{frenkel}
\begin{equation}
\langle\delta\psi(t)|i\partial_{t}-H|\psi(t)\rangle=0,\label{eq:DFVP}
\end{equation}
where $\langle\delta\psi(t)|$ is the variation of the trial state. Hamiltonian \eqref{eq:Ham} can be seen as a variant of the spin-boson model, where the cavity mode can be treated as a special mode of the reservoir. Therefore, the multi-$D_1$~\cite{Yan_2020,Wang_2016,Deng_2016} and multi-$D_2$~\cite{Wang_2021,Wang_2017,Fujihashi_2017,Werther_2019} trial states can be applied to it. It has been pointed out that the former with the multiplicity $M$ can be viewed as a special case of the latter with the multiplicity $2M$~\cite{Werther_2019}. Nevertheless, for the same multiplicity $M$, the multi-$D_1$ state has more variational parameters and may be more feasible to reach convergence than the multi-$D_2$ state. Thus, the multi-$D_1$ ansatz will be used in this work, which reads
\begin{equation}
|D_{1}^{M}(t)\rangle=\sum_{n=1}^{M}\left[A_{n}|+\rangle|f_{n}\rangle+B_{n}|-\rangle|g_{n}\rangle\right],
\end{equation}
where $|\pm\rangle$ are the eigenstates of $\sigma_{x}$ with eigenvalues
$\pm1$. $|f_{n}\rangle$ and $|g_{n}\rangle$ are the multimode coherent
states:
\begin{equation}
|f_{n}\rangle=e^{-\sum_{j}|f_{nj}|^{2}/2}e^{f_{nc}b_{c}^{\dagger}}|0_{c}\rangle\otimes e^{\sum_{k}f_{nk}b_{k}^{\dagger}}|0\rangle,
\end{equation}
\begin{equation}
|g_{n}\rangle=e^{-\sum_{j}|g_{nj}|^{2}/2}e^{g_{nc}b_{c}^{\dagger}}|0_{c}\rangle\otimes e^{\sum_{k}g_{nk}b_{k}^{\dagger}}|0\rangle,
\end{equation}
where $|0_{c}\rangle$ and $|0\rangle$ are the vacuum states
for the cavity and reservoir, respectively, and the subscript $j$ denotes either the cavity or reservoir mode. Here, $A_{n}$, $B_{n}$, $f_{nj}$,
and $g_{nj}$ ($j=c$ or $k$) are complex variational parameters and are
time-dependent functions. The equations of motion for the variational parameters
can be readily derived from \eqref{eq:DFVP}, which are presented in the Supporting Information.

The numerical simulation can be easily carried out in the present formalism. First, the equations of motion are numerically solved to obtain the time derivatives of the variational parameters, $\dot{A}_n$, $\dot{B}_n$, $\dot{f}_{nj}$, and $\dot{g}_{nj}$. Second, the derivatives are used to update the variational
parameters based on the 4th-order Runge-Kutta method. Third, after the integration, the physical quantities of the qubit, cavity, and reservoir can be calculated simultaneously. Specifically,
the photon number of the mode $j$ (either cavity mode or reservoir
mode) can be obtained as
\begin{eqnarray}
N(\omega_{j},t) & = & \langle D_{1}^{M}(t)|b_{j}^{\dagger}b_{j}|D_{1}^{M}(t)\rangle\nonumber \\
 & = & \sum_{l,n}\left(A_{l}^{\ast}f_{lj}^{\ast}S_{ln}^{(f,f)}f_{nj}A_{n}+B_{l}^{\ast}g_{lj}^{\ast}S_{ln}^{(g,g)}g_{nj}B_{n}\right).\nonumber\\
\end{eqnarray}
It is obvious that $N(\omega_j,t)$ as a function of $t$ for a given $\omega_j$ describes the time evolution of the photon number at the $j$ mode; $N(\omega,t)$ as a function of $\omega$ (excluding the cavity mode) at a fixed time $t$ describes the photon number distribution over the frequency of the radiation reservoir.

It is feasible to measure the accuracy of the variational results. Following Refs.~\cite{Yan_2020,Martinazzo_2020}, we calculate the
squared norm of the deviation vector scaled against $\omega_0^2$,
\begin{eqnarray}
\sigma^{2}(t)& = &\left|(i\partial_{t}-H)|D_{1}^{M}(t)\rangle\right|^{2}/\omega_{0}^{2}\nonumber \\
 & = & \omega_{0}^{-2}\left[\left\langle H^{2}\right\rangle -\langle\dot{D}_{1}^{M}(t)|\dot{D}_{1}^{M}(t)\rangle\right],
\end{eqnarray}
where the detailed expression is given in the Supporting Information.
In our previous work, we have shown that the variational approach yields accurate results that are in excellent agreement with the hierarchy equations of motion as long as $\sigma^2(t)<10^{-2}$~\cite{Yan_2020}.
It is therefore possible to acquire the validity of the variational results by monitoring the magnitude of the deviation $\sigma^2(t)$.

Since we are interested in the spontaneous emission spectrum, the initial state of the total system is set as $|e,0_c,0\rangle=\frac{1}{\sqrt{2}}(|+\rangle+|-\rangle)\otimes|0\rangle_{c}\otimes|0\rangle$, namely, the qubit is in excited state and both the cavity and reservoir are in the vacuum states. To perform numerical simulation, we discretize the spectral density function with the nonuniform intervals in $[0,\omega_{\rm max}]$ as in Refs.~\cite{Makri_1999,Wang_2001},
which results in the discretized frequencies and coupling constants:
\begin{equation}
\omega_{k}=-\omega_{{\rm cut}}\ln\left[1-\frac{k}{N_{b}}(1-e^{-\omega_{{\rm max}}/\omega_{{\rm cut}}})\right],
\end{equation}
\begin{equation}
\lambda_{k}=\sqrt{2\alpha\omega_{k}\omega_{{\rm cut}}(1-e^{-\omega_{{\rm max}}/\omega_{{\rm cut}}})/N_{b}},
\end{equation}
where $N_{b}$ is the number of discretized modes. Throughout this work, the cut frequency of the spectral density is set as $\omega_{\rm cut}=5\omega_0$. $\omega_{\rm max}=4\omega_{\rm cut}$ and $N_b=500$ are used for the discretization of the spectral density. We have carefully tested the convergence of the numerical results presented below. The used parameters are sufficient for the discretization of the spectral density to guarantee the convergence of the numerical results provided the multiplicity $M$ is large enough. Moreover, we have examined the deviation of the variational results from the true solutions by monitoring $\sigma^2(t)$. In general, $\sigma^2(t)$ varies with time. Consequently, we single out its maximal magnitude in the time interval $[0,t_f]$ of the evolution to characterize the accuracy. Throughout this work, we set $t_f=300\omega_0^{-1}$. In Table~\ref{tab1}, we show the maximal values of the deviation for various coupling parameters under study. We note that the magnitude of $\sigma^2(t)$ is of the order $10^{-3}$. This suggests that the present variational results are numerically accurate and thus we use them as the benchmark for comparison with analytical results below.

As the reservoir is initially in the vacuum state, the emission spectrum can be defined as the photon number distribution over the frequency of the reservoir mode, which can be given by $N(\omega_k,t)$ as a function of $\omega_k$ at given $t$. Clearly, the defined spectrum is naturally time dependent. To obtain the steady-state emission  spectrum, one needs to propagate the equations of motion in a sufficiently long time. In practice, we find that the final time of the evolution $t_f$ is sufficient for the total system to reach its steady state in most cases when the qubit-cavity and qubit-reservoir couplings are moderately strong, e.g., $\lambda_c/\omega_c>0.1$ and $\alpha\sim0.1$. In the following, we just show $N(\omega_k,t_f)$ and it will be referred to as the multi-$D_1$ result.

\begin{table}
\caption{The maximum values of the deviation $\sigma^{2}(t)$ within the time
interval $[0,t_f]$ for $\omega_c=\omega_0$ and various values of $\lambda_{c}/\omega_{c}$
and $\alpha$. The total number of the reservoir mode is $N_{b}=500$.
The number in the brackets is the multiplicity $M$.}\label{tab1}
\begin{tabular}{cccc}
\hline
$\lambda_{c}/\omega_{c}$ & $\alpha=0.05$ & $\alpha=0.1$ & $\alpha=0.2$\tabularnewline

\hline
$0.0$ & $0.0010\,[3]$ & $0.0016\,[6]$ & $0.0015\,[10]$\tabularnewline

$0.1$ & $0.0010\,[3]$ & $0.0015\,[6]$ & $0.0011\,[12]$\tabularnewline

$0.3$ & $0.0018\,[4]$ & $0.0044\,[6]$ & $0.0023\,[12]$\tabularnewline

$0.5$ & $0.0041\,[4]$ & $0.0047\,[10]$ & $0.0043\,[12]$\tabularnewline
\hline
\end{tabular}
\end{table}

\begin{figure}
  \centering
  \includegraphics[width=\columnwidth]{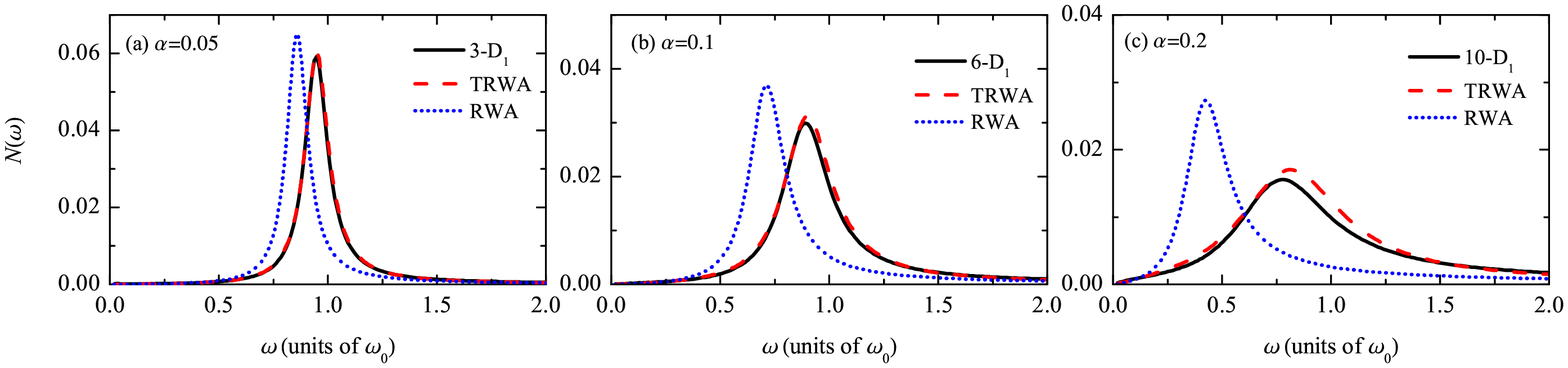}
  \caption{The emission spectra computed from the three methods for $\lambda_c=0$ and the three values of $\alpha$.}\label{fig1}
\end{figure}

\begin{figure}
  \centering
  \includegraphics[width=\columnwidth]{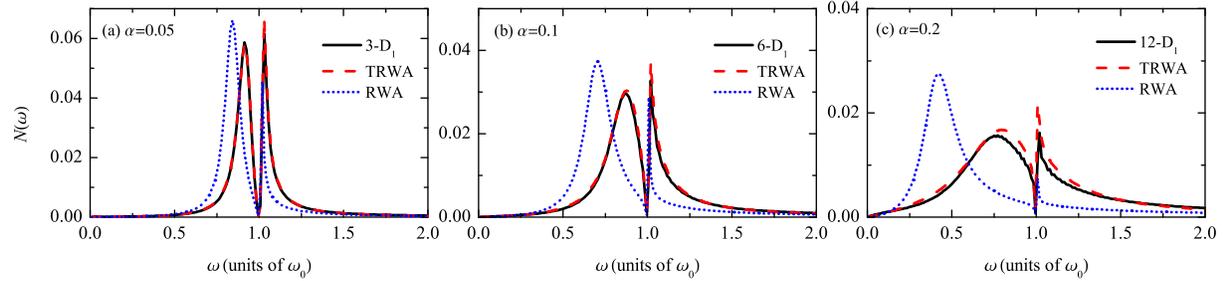}
  \caption{The emission spectra computed from the three methods for $\omega_c=\omega_0$, $\lambda_c=0.1\omega_0$, and the three values of $\alpha$.}\label{fig2}
\end{figure}

In order to give further insights into the influence of the Lamb shift on the vacuum Rabi splitting, we analytically calculate the emission spectrum based on the resolvent formalism~\cite{Cohen} and two different approximations.

The first analytical result relies on a unitary transformation~\cite{Zheng_2004,Gan_2010}, with which we eliminate the counter-rotating coupling in the transformed frame with the reasonable approximation that neglects higher-order coupling terms and thus obtain the resulting effective Hamiltonian with the rotating-wave couplings only. Nevertheless, the effects of the counter-rotating couplings are taken into account with the renormalized quantities. In the steady-state limit,  we find the emission spectrum as follows:
\begin{eqnarray}
N(\omega_{k})
 & = & \left|\frac{\tilde{\lambda}_{k}\left[\omega_{k}-\omega_{c}+\tilde{\lambda}_{c}^{2}/(2\eta\omega_{0})\right]}{(\omega_{k}-\omega_{c})\left[\omega_{k}-\eta\omega_{0}-\tilde{\Delta}(\omega_{k})+i\tilde{\Gamma}(\omega_{k})\right]-\tilde{\lambda}_{c}^{2}}\right.\nonumber\\
 &    &\left.+\frac{\lambda_{k}}{2(\omega_{k}+\eta\omega_{0})}\right|^{2}, \label{eq:swtrwa}
\end{eqnarray}
where
\begin{equation}
\tilde{\lambda}_j=\frac{\eta\omega_{0}\lambda_{j}}{\eta\omega_{0}+\omega_{j}},\quad(j=c,\,k),
\end{equation}
\begin{equation}
 \tilde{\Delta}(\omega) = P\int_{0}^{\infty}\left(\frac{\eta\omega_{0}}{\eta\omega_{0}+x}\right)^{2}\frac{J(x)dx}{(\omega-x)},\label{eq:Dtw}
\end{equation}
\begin{equation}
  \tilde{\Gamma}(\omega)=\pi\left(\frac{\eta\omega_{0}}{\eta\omega_{0}+\omega}\right)^{2}J(\omega),\label{eq:Gtw}
\end{equation}
and the renormalization parameter $\eta$ should be consistently solved from
\begin{equation}
\eta=\exp\left[-\frac{1}{2}\left(\frac{\lambda_{c}}{\omega_{c}+\eta\omega_{0}}\right)^{2}-\frac{1}{2}\int_{0}^{\infty}\frac{J(x)dx}{(x+\eta\omega_{0})^{2}}\right].\label{eq:etaeq}
\end{equation}
The detailed derivation of this result is presented in the Supporting Information. In the above expression, we note that the renormalization parameter $\eta$ and $\tilde{\Delta}(\omega)$ are responsible for the level shift.
Henceforth, the emission spectrum computed from Eq.~\eqref{eq:swtrwa} is referred to as the transformed RWA (TRWA) result. The present TRWA spectrum is different from that presented in Ref.~\cite{Cao_2011}, where a similar TRWA treatment has been combined with the master equation approach to derive a nonRWA emission spectrum. The main difference is that the influence of unitary transformation on the initial state has not been taken into account in that work, which leads to poor performance of the analytical result in the strong coupling regime.

The second analytical result is obtained with the aid of the RWA.
In quantum optics, the RWA is widely used for the model under study, which neglects the counter-rotating couplings and yields the RWA Hamiltonian
\begin{eqnarray}
  H_{{\rm RWA}} &=& \frac{1}{2}\omega_{0}\sigma_{z}+\omega_{c}b_{c}^{\dagger}b_{c}+\frac{\lambda_{c}}{2}(b_{c}\sigma_{+}+b_{c}^{\dagger}\sigma_{-})\nonumber \\
   & & +\sum_{k}\omega_{k}b_{k}^{\dagger}b_{k}+\sum_{k}\frac{\lambda_{k}}{2}(b_{k}\sigma_{+}+b_{k}^{\dagger}\sigma_{-}).
\end{eqnarray}
For the RWA Hamiltonian,
we set $H_{0}=\frac{1}{2}\omega_{0}\sigma_{z}+\omega_{c}b_{c}^{\dagger}b_{c}+\sum_{k}\omega_{k}b_{k}^{\dagger}b_{k}$
and $V=\frac{\lambda_{c}}{2}(b_{c}\sigma_{+}+b_{c}^{\dagger}\sigma_{-})+\sum_{k}\frac{\lambda_{k}}{2}(b_{k}\sigma_{+}+b_{k}^{\dagger}\sigma_{-})$.
We can calculate the emission spectrum by using the resolvent formalism. Alternatively, the RWA spectrum be obtained from the TRWA spectrum by replacing the renormalized quantities with the bare ones, i.e., $\tilde{\lambda}_j\rightarrow\lambda_j/2$, and taking the weak coupling limit, $\eta\rightarrow1$, $\tilde{\lambda}_c^2/(2\eta\omega_0)\rightarrow0$, and $\lambda_k/[2(\eta\omega_0+\omega_k)]\rightarrow0$.
The steady-state emission spectrum with the RWA is given by~\cite{Cao_2011}
\begin{equation}
N(\omega_{k})= \frac{\lambda_{k}^{2}/4}{\left[\omega_{k}-\omega_{0}-\Delta(\omega_{k})-\frac{\lambda_{c}^{2}/4}{\left(\omega_{k}-\omega_{c}\right)}\right]^{2}+\Gamma^{2}(\omega_{k})},\label{eq:NWRWA}
\end{equation}
where
\begin{equation}
\Delta(\omega)=P\int_{0}^{\infty}\frac{J(x)dx}{4(\omega-x)},
\end{equation}
\begin{equation}
\Gamma(\omega)=\frac{\pi}{4}J(\omega).
\end{equation}
Here, $\Delta(\omega)$ is responsible for the level shift. The emission spectrum calculated from Eq.~\eqref{eq:NWRWA} is referred to as the RWA result. Equation~\eqref{eq:NWRWA} implies that the emission spectrum becomes zero as long as $\omega_k=\omega_c$ regardless of the cavity-qubit coupling. Similar behavior can be observed from the multi-$D_1$ and TRWA results. This arises from the fact that the loss of the cavity is not taken into account. Nevertheless, those results can be justified when the decay rate of the qubit and the cavity-qubit coupling are much greater than the leakage of the cavity.

\begin{figure}
  \centering
  \includegraphics[width=\columnwidth]{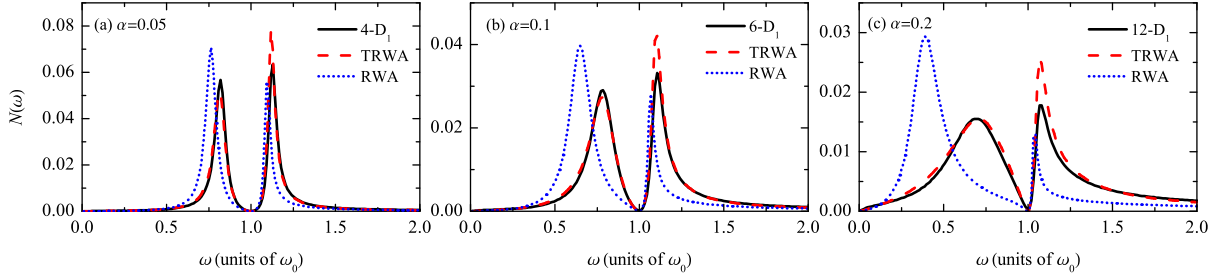}
  \caption{The emission spectra computed from the three methods for $\omega_c=\omega_0$, $\lambda_c=0.3\omega_0$, and the three values of $\alpha$.}\label{fig3}
\end{figure}

\begin{figure}
  \centering
  \includegraphics[width=\columnwidth]{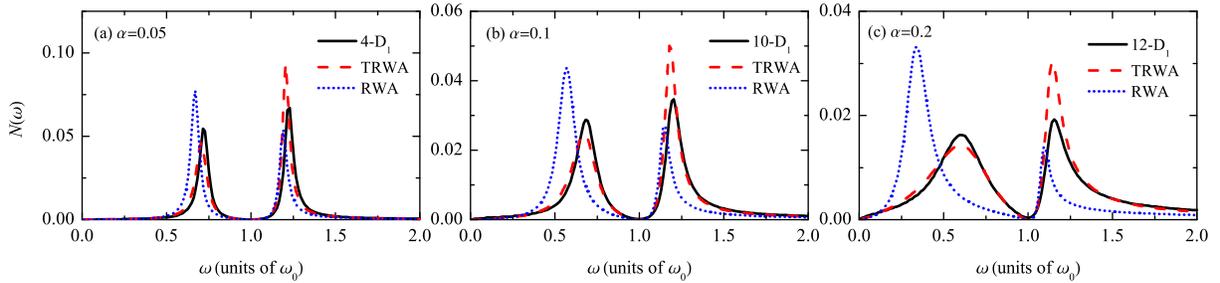}
  \caption{The emission spectra computed from the three methods for $\omega_c=\omega_0$, $\lambda_c=0.5\omega_0$, and the three values of $\alpha$.}\label{fig4}
\end{figure}

To begin with, we study the spontaneous emission spectrum in the absence of the cavity, i.e., $\lambda_c=0$. Figure~\ref{fig1} displays the spectra computed from the three methods for the three values of $\alpha$. Let us first focus on the multi-$D_1$ results (solid lines). We see that the emission lines are peaked on smaller frequencies than the bare transition frequency of the qubit, $\omega_0$. This phenomenon is known as the Lamb shift. From Fig.~\ref{fig1}(a) to~\ref{fig1}(c), it is clear to see that the Lamb shift increases with the increasing of $\alpha$. Besides, the emission line profile becomes different from the standard Lorentzian and the emission peak becomes broader. Moreover, it is worthwhile to note that linewidth of the spectrum is comparable to the bare transition frequency of the qubit, indicating an ultrastrong light-matter coupling regime. When comparing the approximate results with the multi-$D_1$ results, we find that when $\lambda_c=0$, the TRWA treatment (dashed line) is capable to provide a good approximate description of the spontaneous emission even in the strong coupling regime $\alpha=0.2$. On the contrary, comparing the RWA results (dotted lines) with the multi-$D_1$ results, we find that the RWA fails to predict the correct emission spectrum for all the three values of $\alpha$.

It is worthwhile to note that the magnitude of the Lamb shift predicted with the RWA is much larger than that of the multi-$D_1$, namely, the RWA leads to the overestimate of the Lamb shift. To get insights into why the RWA overestimates the Lamb shift, we can examine the main difference between the TRWA and RWA spectra, which arises from the renormalization factor
\begin{equation}
  \left(\frac{\eta\omega_0}{\eta\omega_0+\omega}\right)^2,
\end{equation}
incorporated in Eqs. \eqref{eq:Dtw} and \eqref{eq:Gtw}. Owing to this factor, we find that the TRWA and RWA level shift terms satisfy $0>\tilde{\Delta}(\omega)>\Delta(\omega)$ when $\omega$ is close to $\omega_0$. This leads to that the RWA overestimates the Lamb shift. The physical reason behind this can be understood by considering the origin of renormalization factor in the TRWA theory. The renormalization factor results from the parameters used in the unitary transformation, $\xi_k=\omega_k/(\eta\omega_0+\omega_k)$, which depends on the boson frequency and determines the displacement of the $k$th bath mode~\cite{L__2007}. When $\omega_k\gg\eta\omega_0$, we have $\xi_k\approx1$. Physically, this means that the bath modes can follow the motion of the system.   When $\omega_k\ll\eta\omega_0$, we have $\xi_k\ll1$. Physically, this means that the corresponding bath modes are too slow to follow the motion. However, it turns out that this nature of the bath modes is not captured by the RWA method and leads to the overestimation. On the other hand, the RWA emission lines are narrower than the multi-$D_1$ ones, indicating the underestimate of the linewidth or the emission rate. This also leads to the conclusion that the counter-rotating terms play an essential role in the Lamb shift and emission rate in the ultrastrong light-matter coupling regimes.

Next, we study the spontaneous emission spectrum in the presence of the cavity. Figure~\ref{fig2} shows the emission spectra computed from the three methods for $\omega_c=\omega_0$, $\lambda_c=0.1\omega_0$, and the three values of $\alpha$. For each value of $\alpha$, we find that the emission spectrum has two peaks. Moreover, we note that the left peak is broader than the right peak, which is different from the typical vacuum Rabi splitting profile when the qubit and cavity are on resonance (which consists of two symmetric Lorentzian peaks for the RWA and two Lorentzian peaks different in heights for the nonRWA)~\cite{Auff_ves_2008,Cao_2011}. Besides, the difference in the linewidth of the two peaks becomes larger as the increasing of $\alpha$. This type of spectral feature signifies an off-resonant coupling between the cavity and the qubit and can be attributed to the Lamb shift, which leads to a considerable detuning from the renormalized transition frequency of the qubit to the cavity frequency ($\omega_c=\omega_0$). This assessment is confirmed by using the analytical methods. It turns out that if we remove the Lamb shift in Eqs. \eqref{eq:swtrwa} and \eqref{eq:NWRWA}, the emission spectral profile changes significantly from those presented in Fig.~\ref{fig2} and the feature that the spectrum exhibits a broad peak and a narrow peak vanishes. When comparing the TRWA and RWA results with the multi-$D_1$ results, we find that the TRWA treatment is still capable to give a relatively good approximate description [In Fig.~\ref{fig2}(c), one notes that the multi-$D_1$ curve is not as smooth as other curves, which is because the system does not completely reach its steady state at $t=300\omega_0^{-1}$]. However, the RWA results are found to be apparently different from the multi-$D_1$ results. The RWA spectral profile actually arises from the fact that the RWA leads to the overestimate of the Lamb shift. The large shift leads to a large detuning from the shifted transition frequency of the qubit to the cavity frequency. Consequently, the spontaneous emission of the qubit is weakly influenced by the cavity when the cavity-qubit coupling is not strong enough. The present results show that the increase of the qubit-reservoir coupling leads to the non-negligible Lamb shift, which essentially modifies the vacuum Rabi splitting profile.

To further examine the influence of the Lamb shift on the vacuum Rabi splitting, we compute the emission spectrum by using the three methods for $\omega_c=\omega_0$, $\lambda_c=0.3\omega_0$, and the three values of $\alpha$. In Fig.~\ref{fig3}(a), we see that when $\alpha=0.05$, the TRWA and multi-$D_1$ spectral profiles somewhat resemble the typical vacuum Rabi splitting profile. However, the RWA spectrum still possesses two peaks with significantly different linewidths. This can be understood by noting that the Lamb shift of the RWA is much larger than the other two treatments for $\alpha=0.05$. From Fig.~\ref{fig3}(a) to \ref{fig3}(c), for each treatment, the increase of $\alpha$ leads to the similar modification of the spectral profile, namely, the linewidths of the two emission peaks become apparently different, which is associated with the increase of the Lamb shift. This change of the spectrum with the variation of $\alpha$ is similar to Fig.~\ref{fig2}. In addition, we note that for $\omega_c=\omega_0$ and $\lambda_c=0.3\omega_c$, there is a quantitative discrepancy in the heights of the higher-frequency emission peaks between the TRWA and the multi-$D_1$ results, indicating the inadequacy of the TRWA treatment when the qubit-cavity coupling is sufficiently strong.

In Fig.~\ref{fig4}, we show the emission spectra obtained from the three methods for $\omega_c=\omega_0$, $\lambda_c=0.5\omega_0$, and the three values of $\alpha$. We note that the spectral profiles of each treatment are similar to those presented in Fig.~\ref{fig3}. Nevertheless, when comparing Figs.~\ref{fig2}(a), \ref{fig3}(a), and \ref{fig4}(a), we note that for a fixed $\alpha$, as the increasing of $\lambda_c$, the low-frequency peak narrows while the high frequency peak broadens. Similar situation also occurs for $\alpha=0.1$ and $0.2$. This reflects the modification of the spontaneous emission induced by the ultrastrong strong qubit-cavity interaction even if there is a considerable detuning caused by the Lamb shift.  To gain a comprehensive understanding of the transformation of the spectrum with the variation of the cavity-qubit coupling, one needs to analyze energies and decay rates of the two polariton states associated with the two emission peaks, which can be derived from the poles of the resolvent operator~\cite{Cohen}. For a relatively small value of $\lambda_c$ the two polariton states can be quite different in decay rates. For a relatively large value of $\lambda_c$, the two polariton states can have similar decay rates. As a result, one observes the transformation of the spectra shown in Figs. \ref{fig2}-\ref{fig4}. We present the analytical calculation of the energies and decay rates of the polarition states with the Markovian approximation in the Supporting Information for the qualitative explanation of the spectral feature. In addition, Fig.~\ref{fig4} suggests that the larger the qubit-cavity coupling is, the larger the discrepancy between the TRWA and the multi-$D_1$ methods becomes. This further confirms that the TRWA treatment is inadequate if the qubit-cavity coupling is strong enough. Nevertheless, it can be expected that the TRWA treatment is valid when $\lambda_c/\omega_0\sim0.1$, $\omega_c/\omega_0\sim1$, and $\alpha<0.2$. On the contrary, Figs.~\ref{fig1}-\ref{fig4} show that the RWA completely breaks down in this regime of the parameters.

In summary, we have studied the spontaneous emission of the qubit coupled simultaneously to a high-$Q$ cavity and a radiation reservoir in the regimes of ultrastrong light-matter coupling by using the numerically eaxct variational approach and two analytical approaches based on the TRWA and the RWA. It has been demonstrated that the Dirac-Frenkel variational principle can be characterized by the squared norm of the deviation vector. The TRWA treatment is found to provide good approximate results in comparison with those from the multi-$D_1$ method if $\lambda_c\sim0.1\omega_0$, $\omega_c\sim\omega_0$, and $\alpha<0.2$. The RWA treatment, however, fails completely in the parameter regime under study. In addition, contrasting the RWA results with non-RWA ones reveals that the RWA overestimates (underestimates) the Lamb shift (emission rate). In comparison, our variational approach presented here can be easily extended to cavity-molecular systems and polariton chemistry.

Combining the three aforementioned approaches, we have illustrated in this work that the Lamb shift plays an important role in modifying the vacuum Rabi splitting profile in the ultrastrong coupling regimes. If $\omega_c=\omega_0$, the emission spectrum exhibits two peaks with apparently different linewidth and height for a relatively small cavity-qubit coupling strength, a signature of the off-resonant qubit-cavity coupling that deviates significantly from the typical vacuum Rabi splitting profile for the resonant qubit-cavity coupling. As the cavity-qubit coupling increases,  the low-frequency peak narrows and the high-frequency peak broadens until they have similar widths. Our finding suggests that the Lamb shift should be properly taken into account when studying the resonant interaction between a strongly dissipative qubit and a cavity.

\begin{acknowledgement}
Support from the National Natural Science Foundation of China (Grants No. 12005188, No. 11774226, and No. 11774311)
and the Singapore Ministry of Education Academic Research
Fund Tier 1 (Grant No. RG190/18) is gratefully acknowledged.
\end{acknowledgement}

\begin{suppinfo}

The following file is available free of charge.
\begin{itemize}
  \item SI\_VRS\_v7: Equations of motion for the variational parameters, brief review of the resolvent formalism, analytical derivation of the emission spectrum, transformed rotating-wave approximation, energies and decay rates of the polariton states.

\end{itemize}

\end{suppinfo}

\bibliography{VRSbib}

\end{document}